# Robust Stabilization of Fractional-order Interval Systems via Dynamic Output Feedback: An LMI Approach

Pouya Badri[1], Mahdi Sojoodi[1*]

[1]Advanced Control Systems Laboratory, School of Electrical and Computer Engineering, Tarbiat Modares University, Tehran, Iran.
[*]sojoodi@modares.ac.ir

**Abstract:** This paper addresses the problem of robust dynamic output stabilization of FO-LTI interval systems with the fractional order $0 < \alpha < 2$, in terms of linear matrix inequalities (LMIs). Our purpose is to design a robust dynamic output feedback controller that asymptotically stabilizes interval fractional-order linear time-invariant (FO-LTI) systems. Sufficient conditions are obtained for designing a stabilizing controller with a predetermined order, which can be chosen to be as low as possible. The LMI-based procedures of designing robust stabilizing controllers are preserved in spite of the complexity of assuming the most complete model of linear controller, with direct feedthrough parameter. Finally, some numerical examples with simulations are presented to demonstrate the effectiveness and correctness of the theoretical results.

## 1. Introduction

In recent years fractional-order control systems have attracted increasing interest and attention of physicists and engineers from an application viewpoint [1]–[4]. Fractional-order models can more concisely and actually describe systems having responses with long memory transients. Furthermore, it has long been recognized that a large number of natural and biological systems have intrinsic features that can be better described by fractional-order models [5]–[9]. Hence, controller-designing problems for fractional-order systems are among the most interesting issues in the literature. Using fractional-order controllers is a suitable method for controlling such systems, since it has been demonstrated that fractional-order controllers can capture much better effect and robustness [10], [11]. Since, within the most of control systems, stability is the primary objective to be accomplished, and due to uncertain models caused by neglected dynamics, parametric variations in time, uncertain physical parameters, and so on, robust stability and stabilization problems became a basic issue for all control systems as well as fractional-order systems [12]–[15].

In [12]–[17], robust stability and stabilization analysis on fractional-order control systems were presented. Necessary and sufficient conditions for the stability and stabilization of fractional-order interval systems are investigated in [12]. Moreover, for the robust stability of general interval fractional-order system in which interval uncertainties exist both in the coefficients and orders of the FO-LTI system, necessary and sufficient conditions are presented in [18]. In [13] the robust stability and stabilization of fractional-order linear systems with positive real uncertainty are presented, where the existence conditions and design procedures of the static state feedback controller, static output feedback controller and observer-based controller for asymptotically stabilizing of such systems are investigated with the constraint on the output matrix to be of full-row rank.

In the most of mentioned methods, state feedback controller is utilized, in which all individual states must be accessible. Nevertheless, in most practical applications measuring all states is impossible owing to economic issues or physical constraints [19], where employing output feedback control is useful. It worth mentioning that dynamic feedback controller brings about more effective control performances, more flexibility, and more degrees of freedom in achieving control objectives, compared to the static controller [20]. Furthermore, it is easy to show that some unstable systems cannot be stabilized using static controllers and their stabilization entails using dynamic controllers [21]. Therefore, we have proposed a robust dynamic output feedback controller for FO-LTI systems with positive real uncertainty by means of linear matrix inequalities in [15]. On the other hand, it has been admitted that system modeling with parametric interval uncertainty is more convenient for the control system design problems [22] and also is more suitable for robust stability analysis [23].

In [24], observer-based robust stabilization of a class of Lipschitz non-linear fractional-order systems is addressed, where stabilization of fractional-order interval systems with fractional order $0 < \alpha < 1$ is possible if the output matrix of the uncertain system is of full row rank. Moreover, stabilization of fractional-order systems, with fractional order $0 < \alpha < 1$, subject to bounded uncertainties is investigated in [25], where the system uncertainties are randomly distributed in the state matrix $A$ and the output matrix $C$. In this paper, it is assumed that all individual possible pairs of $A + \Delta A$, $C + \Delta C$ are observable in the sense of Kalman. Furthermore, in [26] by using singular value decomposition and linear matrix inequality techniques, robust control of fractional-order interval linear systems with fractional order $1 \leq \alpha < 2$ was considered by assuming that the output matrix of the uncertain system is of full row rank.

In the majority of available controller design methods, high-order controllers are obtained suffering from costly implementation, unfavorable reliability, high fragility, maintenance difficulties, and potential numerical errors. Designing a controller with a low and fixed order would be helpful, since the closed-loop performance is not necessarily guaranteed by available plant or controller order reduction methods. [19].

Motivated by aforementioned observations, our paper aims at solving the problem of robust dynamic output stabilization of fractional-order linear interval systems with the fractional order $0 < \alpha < 2$, using a fixed-order dynamic output feedback controller in terms of linear matrix inequalities (LMIs). In spite of the intricacy of considering the most complete model of linear controller with direct feedthrough parameter, LMI approach of designing robust control is preserved, which is suitable to be used in practice due to various efficient convex optimization parsers and solvers that can be used to determine the feasibility of the LMI constraints and calculate design parameters. Furthermore, in our proposed method designing a dynamic feedback controller does not lead to limiting constraints on the state space matrices of the uncertain systems which occur in previous works.

As far as we know, there is no result on the analytical design of a stabilizing fixed-order dynamic output feedback controller for fractional-order systems with interval uncertainties in the literature. In this paper, sufficient conditions are obtained for designing a robust stabilizing controller with a predetermined order, which can be chosen to be as low as possible for simpler implementation.

The rest of this paper is organized as follows: In section 2, some preliminaries about interval uncertainty and fractional-order calculus together with the problem formulation are presented. LMI-based robust stabilizing conditions of fractional-order interval systems using a dynamic output feedback controller are derived in Section 3. Some numerical examples are given in Section 4 to demonstrate the effectiveness of the proposed theoretical results. Finally, conclusion is drawn in section 5.

**Notations**: In this paper $A \otimes B$ denotes the kronecker product of matrices $A$ and $B$ and by $M^T$, $\overline{M}$ and $M^*$, we denote the transpose, the conjugate, and the transpose conjugate of $M$, resepectively. The conjugate of the scalar number $z$ is represented by $\bar{z}$ and $Sym(M)$ denotes $M + M^*$. The notation $\bullet$ is the symmetric component symbol in matrix and $\uparrow$ is the symbol of pseudo inverse. The notations $\mathbf{0}$ and $I$ denote the zero and identity matrices with appropriate dimensions and $i$ stands for the imaginary unit.

## 2. Preliminaries and problem formulation

In this section, some basic concepts and lemmas of fractional-order calculus and interval uncertainty are presented.

Consider the following uncertain FO-LTI system:

$$D^\alpha x(t) = Ax(t) + Bu(t), 0 < \alpha < 2 \qquad (1)$$
$$y(t) = Cx(t)$$

in which $x \in R^n$ denotes the pseudo-state vector, $u \in R^l$ is the control input, and $y \in R^m$ is the output vector. Furthermore, $A \in R^{n\times n}$ and $B \in R^{n\times l}$ are interval uncertain matrices as follows

$$A \in A_I = [\underline{A}, \overline{A}] = \{[a_{ij}]: \underline{a}_{ij} \leq a_{ij} \leq \overline{a}_{ij}, 1 \leq i,j \leq n\}, \qquad (2)$$

$$B \in B_I = [\underline{B}, \overline{B}] = \{[b_{ij}]: \underline{b}_{ij} \leq b_{ij} \leq \overline{b}_{ij}, 1 \leq i \leq n, 1 \leq j \leq l\}, \qquad (3)$$

where $\underline{A} = [\underline{a}_{ij}]_{n\times n}$ and $\overline{A} = [\overline{a}_{ij}]_{n\times n}$ satisfy $\underline{a}_{ij} \leq \overline{a}_{ij}$ for all $1 \leq i,j \leq n$, $\underline{B} = [\underline{b}_{ij}]_{n\times l}$ and $\underline{B} = [\underline{b}_{ij}]_{n\times l}$ satisfy $\underline{b}_{ij} \leq \overline{b}_{ij}$ for all $1 \leq i \leq n, 1 \leq j \leq l$.

In this paper, the following Caputo definition for fractional derivatives of order α of function $f(t)$ is adopted since initial values of classical integer-order derivatives with clear physical interpretations are utilizable using the Laplace transform of the Caputo derivative [27]:

$$_a^C D_t^\alpha f(t) = \frac{1}{\Gamma(m-\alpha)} \int_a^t (t-\tau)^{m-a-1} \left(\frac{d}{d\tau}\right)^m f(\tau) d\tau$$

where $\Gamma(\cdot)$ is Gamma function defined by $\Gamma(\epsilon) = \int_0^\infty e^{-t} t^{\epsilon-1} dt$ and $m$ is the smallest integer that is equal to or greater than $\alpha$.

The following notations are introduced in order to deal with the interval uncertainties

$$A_0 = 1/2(\underline{A} + \overline{A}), \Delta A = 1/2(\overline{A} - \underline{A}) = \{\gamma_{ij}\}_{n\times n}, \qquad (4)$$

$$B_0 = 1/2(\underline{B} + \overline{B}), \Delta B = 1/2(\overline{B} - \underline{B}) = \{\beta_{ij}\}_{n\times l}, \qquad (5)$$

It is obvious that all elements of ΔA and ΔB are nonnegative, so the following matrices can be defined

$$M_A = [\sqrt{\gamma_{11}}e_1^n \quad \ldots \quad \sqrt{\gamma_{1n}}e_1^n \quad \ldots \quad \sqrt{\gamma_{n1}}e_n^n \quad \ldots \quad \sqrt{\gamma_{nn}}e_n^n]_{n\times n^2}, \qquad (6)$$

$$R_A = [\sqrt{\gamma_{11}}e_1^n \quad \ldots \quad \sqrt{\gamma_{1n}}e_n^n \ldots \quad \sqrt{\gamma_{n1}}e_1^n \quad \ldots \quad \sqrt{\gamma_{nn}}e_n^n]_{n^2\times n}^T, \qquad (7)$$

$$M_B = [\sqrt{\beta_{11}}e_1^n \quad \ldots \quad \sqrt{\beta_{1l}}e_1^n \quad \ldots \sqrt{\beta_{n1}}e_n^n \quad \ldots \quad \sqrt{\beta_{nl}}e_n^n]_{n\times nl}, \qquad (8)$$

$$R_B = [\sqrt{\beta_{11}}e_1^l \quad \ldots \quad \sqrt{\beta_{1l}}e_l^l \quad \ldots \sqrt{\beta_{n1}}e_1^l \quad \ldots \quad \sqrt{\beta_{nl}}e_l^l]_{nl\times l}^T, \qquad (9)$$

where $e_k^n \in R^n$, $e_k^l \in R^l$, and $e_k^m \in R^m$ are column vectors with the $k$-th element being 1 and all the others being 0. Also, we have

$$H_A = \{diag(\delta_{11}, \ldots, \delta_{1n}, \ldots, \delta_{n1}, \ldots, \delta_{nn}) \in R^{n^2\times n^2}, |\delta_{ij}| \leq 1, i,j \; 1, \ldots, n\}, \qquad (10)$$

$$H_B = \{diag(\eta_{11}, \ldots, \eta_{1l}, \ldots, \eta_{n1}, \ldots, \eta_{nl}) \in R^{(nl) \times (nl)}, |\eta_{ij}| \leq 1, i = 1, \ldots, n, j = 1, \ldots, l\}, \quad (11)$$

In order to study the stability of fractional-order systems and obtain main results the following lemmas are required.

**Lemma 1** [12]: Let

$$A_J = \{A = A_0 + M_A F_A R_A | F_A \in H_A\}, B_J = \{B = B_0 + M_B F_B R_B | F_B \in H_B\}, \quad (12)$$

then $A_I = A_J$, and $B_I = B_J$.

**Lemma 2** [28]: Let $A \in R^{n \times n}$, $0 < \alpha < 1$ and $\theta = (1-\alpha)\pi/2$. The fractional-order system $D^\alpha x(t) = Ax(t)$ is asymptotically stable if and only if there exists a positive definite Hermitian matrix $X = X^* > 0$, $X \in C^{n \times n}$ such that

$$(rX + \bar{r}\bar{X})^T A^T + A(rX + \bar{r}\bar{X}) < 0, \quad (13)$$

where $r = e^{\theta i}$.

**Lemma 3** [29]: Let $A \in \mathcal{R}^{n \times n}$, $1 \leq \alpha < 2$ and $\theta = \pi - \alpha\pi/2$. The fractional-order system $D^\alpha x(t) = Ax(t)$ is asymptotically stable if and only if there exists a positive definite matrix $X \in \mathcal{R}^{n \times n}$ such that

$$\begin{bmatrix} (A^T X + XA)\sin\theta & (XA - A^T X)\cos\theta \\ \bullet & (A^T X + XA)\sin\theta \end{bmatrix} < 0, \quad (14)$$

defining

$$\Theta = \begin{bmatrix} \sin\theta & -\cos\theta \\ \cos\theta & \sin\theta \end{bmatrix}, \quad (15)$$

and with this in mind that $A$ is similar to $A^T$, inequality (15) can be expressed as follows

$$Sym\{\Theta \otimes (AX)\} < 0. \quad (16)$$

**Lemma 4** [12]: For any matrices $X$ and $Y$ with appropriate dimensions, we have

$$X^T Y + Y^T X \leq \eta X^T X + (1/\eta) Y^T Y \text{ for any } \eta > 0. \quad (17)$$

## 3. Main results

The main objective of this paper is to design a robust dynamic output feedback controller that asymptotically stabilizes the interval FO-LTI system (1) in terms of linear matrix inequalities (LMIs). Therefore, the following dynamic output feedback controller is presented

$$\begin{aligned} D^\alpha x_C(t) &= A_C x_C(t) + B_C y(t), \quad 0 < \alpha < 2 \\ u(t) &= C_C x_C(t) + D_C y(t), \end{aligned} \quad (18)$$

with $x_C \in \mathcal{R}^{n_c}$, in which $n_c$ is the arbitrary order of the controller and $A_C, B_C, C_C,$ and $D_C$ are corresponding matrices to be designed.

The resulted closed-loop augmented FO-LTI system using (1) and (11) is as follows

$$D^\alpha x_{Cl}(t) = A_{Cl} x_{Cl}(t), \quad 0 < \alpha < 2 \quad (19)$$

where

$$x_{Cl}(t) = \begin{bmatrix} x(t) \\ x_C(t) \end{bmatrix}, \quad A_{Cl} = \begin{bmatrix} A + BD_C C & BC_C \\ B_C C & A_C \end{bmatrix}. \quad (20)$$

**Theorem 1**: Considering closed-loop system in (19), with $0 < \alpha < 1$, $A \in A_I$, $B \in B_I$, and output matrix $C$, together with a positive definite Hermitian matrix $P = P^*$ in the form of

$$P = diag(P_S, P_C), \quad (21)$$

with $P_S \in C^{n \times n}$ and $P_C \in C^{n_c \times n_c}$ and a real scalar constant $\eta > 0$ alongside with matrices $T_i, i = 1, \ldots, 4$ exist such that the following LMI constrain become feasible

$$\begin{bmatrix} \Sigma + \eta MM^T & R^T \\ \bullet & -\eta I \end{bmatrix} < 0, \quad (22)$$

in which

$$\Sigma = \begin{bmatrix} A_0(rP_S + \bar{r}\overline{P_S}) + (rP_S + \bar{r}\overline{P_S})^T A_0^T + B_0 T_4 + T_4^T B_0^T & B_0 T_3 + T_2^T \\ T_3^T B_0^T + T_2 & T_1 + T_1^T \end{bmatrix}, M = \begin{bmatrix} M_A & M_B & 0 \\ 0 & 0 & 0 \end{bmatrix},$$

$$R = \begin{bmatrix} R_A(rP_S + \bar{r}\overline{P_S}) & 0 \\ R_B D_C C(rP_S + \bar{r}\overline{P_S}) & R_B C_C(rP_C + \bar{r}\overline{P_C}) \\ 0 & 0 \end{bmatrix},$$

(23)

where $\theta = (1 - \alpha)\pi/2$ then, the dynamic output feedback controller parameters of

$$A_C = T_1(rP_C + \bar{r}\overline{P_C})^{-1}, B_C = T_2(rP_S + \bar{r}\overline{P_S})^{-1}C^\uparrow, C_C = T_3(rP_C + \bar{r}\overline{P_C})^{-1}, D_C = T_4(rP_S + \bar{r}\overline{P_S})^{-1}C^\uparrow,$$

(24)

make the closed-loop system in (19) asymptotically stable.

**Proof**: Let $A_{Cl} = A_{0Cl} + A_{\Delta Cl}$, with

$$A_{0Cl} = \begin{bmatrix} A_0 + B_0 D_C C & B_0 C_C \\ B_C C & A_C \end{bmatrix}, A_{\Delta Cl} = \begin{bmatrix} M_A F_A R_A + M_B F_B R_B D_C C & M_B F_B R_B C_C \\ 0 & 0 \end{bmatrix},$$

(25)

It follows from Lemma 2 that the uncertain fractional-order closed-loop system (19) with $0 < \alpha < 1$ is asymptotically stable if there exists a positive definite Hermitian matrix $P = P^*, P \in \mathcal{C}^{(n+n_C) \times (n+n_C)}$ such that

$$(rP + \bar{r}\overline{P})^T A_{Cl}^T + A_{Cl}(rP + \bar{r}\overline{P}) < 0 \Leftrightarrow$$
$$\begin{bmatrix} \Pi_{11} & \Pi_{12} \\ \Pi_{21} & \Pi_{22} \end{bmatrix} + Sym \left\{ \begin{pmatrix} (M_A F_A R_A + M_B F_B R_B D_C C)(rP_S + \bar{r}\overline{P_S}) & M_B F_B R_B C_C(rP_C + \bar{r}\overline{P_C}) \\ 0 & 0 \end{pmatrix} \right\} < 0$$

(26)

in which

$$\Pi_{11} = A_0(rP_S + \bar{r}\overline{P_S}) + (rP_S + \bar{r}\overline{P_S})^T A_0^T + B_0 D_C C P_S + P_S^T C^T D_C^T B_0^T, \Pi_{12} =$$
$$B_0 C_C(rP_C + \bar{r}\overline{P_C}) + (rP_S + \bar{r}\overline{P_S})^T C^T B_C^T, \Pi_{21} = B_C C(rP_S + \bar{r}\overline{P_S}) + (rP_C + \bar{r}\overline{P_C})^T C_C^T B_0^T, \Pi_{22} = A_C P_C + P_C A_C^T.$$

(27)

Applying Lemma 4 to the second part in the right side of (26), one has

$$Sym \left\{ \begin{pmatrix} M_A F_A R_A(rP_S + \bar{r}\overline{P_S}) + M_B F_B R_B D_C C P_S & M_B F_B R_B C_C(rP_C + \bar{r}\overline{P_C}) \\ 0 & 0 \end{pmatrix} \right\}$$
$$= Sym \left\{ \begin{bmatrix} M_A & M_B & 0 \\ 0 & 0 & 0 \end{bmatrix} \times \begin{bmatrix} F_A & 0 & 0 \\ 0 & F_B & 0 \\ 0 & 0 & 0 \end{bmatrix} \times \begin{bmatrix} R_A(rP_S + \bar{r}\overline{P_S}) & 0 \\ R_B D_C C(rP_S + \bar{r}\overline{P_S}) & R_B C_C(rP_C + \bar{r}\overline{P_C}) \\ 0 & 0 \end{bmatrix} \right\}$$
$$\leq \gamma^{-1} \begin{bmatrix} R_A(rP_S + \bar{r}\overline{P_S}) & 0 \\ R_B D_C C(rP_S + \bar{r}\overline{P_S}) & R_B C_C(rP_C + \bar{r}\overline{P_C}) \\ 0 & 0 \end{bmatrix}^T \times \begin{bmatrix} R_A(rP_S + \bar{r}\overline{P_S}) & 0 \\ R_B D_C C(rP_S + \bar{r}\overline{P_S}) & R_B C_C(rP_C + \bar{r}\overline{P_C}) \\ 0 & 0 \end{bmatrix}$$
$$+ \gamma \begin{bmatrix} M_A & M_B & 0 \\ 0 & 0 & 0 \end{bmatrix} \times \begin{bmatrix} M_A & M_B & 0 \\ 0 & 0 & 0 \end{bmatrix}^T.$$

(28)

Substituting (28) into (26) and using the Schur complement of (22) one has

$$\begin{bmatrix} \hat{\Sigma} + \eta M M^T & R^T \\ \bullet & -\eta I \end{bmatrix} < 0,$$

(29)

in which

$$\hat{\Sigma} = \begin{bmatrix} \hat{\Sigma}_{11} & \hat{\Sigma}_{12} \\ \hat{\Sigma}_{21} & \hat{\Sigma}_{22} \end{bmatrix}, \hat{\Sigma}_{11} = A(rP_S + \bar{r}\overline{P_S}) + (rP_S + \bar{r}\overline{P_S})^T A^T + B_0 D_C C P_S + P_S^T C^T D_C^T B_0^T,,$$
$$\hat{\Sigma}_{12} = B C_C(rP_C + \bar{r}\overline{P_C}) + (rP_S + \bar{r}\overline{P_S})^T C^T B_C^T, \hat{\Sigma}_{21} = B_C C(rP_S + \bar{r}\overline{P_S}) + (rP_C + \bar{r}\overline{P_C})^T C_C^T B^T,$$
$$\hat{\Sigma}_{22} = A_C(rP_C + \bar{r}\overline{P_C}) + (rP_C + \bar{r}\overline{P_C})^T A_C^T.$$

(30)

The matrix inequality (36) is not linear because of various multiplications of variables. Accordingly, by changing variables as follows

$$T_1 = A_C(rP_C + \bar{r}\overline{P_C}), T_2 = B_C C(rP_S + \bar{r}\overline{P_S}), T_3 = C_C(rP_C + \bar{r}\overline{P_C}), T_4 = D_C C(rP_S + \bar{r}\overline{P_S}),$$

(31)

the matrix inequality (22) is obtained. ∎

**Theorem 2**: Considering closed-loop system in (19) with $1 \leq \alpha < 2$, $A \in A_I$, $B \in B_I$, and certain output matrix $C$, together with a positive definite symmetric matrix $P = P^T$ in the form of (21) with $P_S \in \mathcal{R}^{n \times n}$ and $P_C \in \mathcal{R}^{n_c \times n_c}$ and a real scalar constant $\eta > 0$ alongside with matrices $T_i, i = 1, \dots, 4$ exist such that the following LMI constrain become feasible

$$\begin{bmatrix} \Sigma + \eta MM^T & R^T \\ \bullet & -\eta I \end{bmatrix} < 0, \tag{32}$$

in which

$$\Sigma = \begin{bmatrix} \Sigma_{11} & \Sigma_{12} \\ \Sigma_{21} & \Sigma_{22} \end{bmatrix}, \Sigma_{11} = \Sigma_{22} = \begin{bmatrix} A_0 P_S + P_S A_0^T + B_0 T_4 + T_4^T B_0^T & B_0 T_3 + T_2^T \\ T_2 + T_3^T B_0^T & T_1 + T_1^T \end{bmatrix} \sin\theta,$$

$$\Sigma_{12} = -\Sigma_{21} = \begin{bmatrix} A_0 P_S - P_S A_0^T + B_0 T_4 - T_4^T B_0^T & B_0 T_3 - T_2^T \\ T_2 - T_3^T B_0^T & T_1 - T_1^T \end{bmatrix} \cos\theta, R = I_2 \otimes \begin{bmatrix} R_A P_S & 0 \\ R_B T_4 & E_B T_3 \\ 0 & 0 \end{bmatrix}, \tag{33}$$

$$M = \begin{bmatrix} M_{11} & M_{12} \\ M_{21} & M_{22} \end{bmatrix}, D_{11} = M_{22} = \begin{bmatrix} M_A & M_B & 0 \\ 0 & 0 & 0 \end{bmatrix} \sin\theta, M_{12} = -M_{21} = \begin{bmatrix} M_A & M_B & 0 \\ 0 & 0 & 0 \end{bmatrix} \cos\theta,$$

where $\theta = \pi - \alpha\pi/2$ then, the dynamic output feedback controller parameters of

$$A_C = T_1 P_C^{-1}, B_C = T_2 P_S^{-1} C^\uparrow, C_C = T_3 P_C^{-1}, D_C = T_4 P_S^{-1} C^\uparrow, \tag{34}$$

make the closed-loop system in (19) asymptotically stable.

**Proof**: Let $A_{Cl} = A_{0Cl} + A_{\Delta Cl}$, which is defined in (25). It follows from Lemma 3 that the uncertain fractional-order closed-loop system (19) with $1 < \alpha \leq 2$ is asymptotically stable if there exists a positive definite matrix $P = P^T$, $P \in \mathcal{R}^{(n+n_C)\times(n+n_C)}$ such that

$$\begin{bmatrix} (A_{Cl}P + PA_{Cl}^T)\sin\theta & (A_{Cl}P - PA_{Cl}^T)\cos\theta \\ (PA_{Cl}^T - A_{Cl}P)\cos\theta & (A_{Cl}P + PA_{Cl}^T)\sin\theta \end{bmatrix}$$
$$= \begin{bmatrix} (A_{0Cl}P + PA_{0Cl}^T)\sin\theta & (A_{0Cl}P - PA_{0Cl}^T)\cos\theta \\ (PA_{0Cl}^T - A_{0Cl}P)\cos\theta & (A_{0Cl}P + PA_{0Cl}^T)\sin\theta \end{bmatrix} \tag{35}$$
$$+ \begin{bmatrix} (A_{\Delta Cl}P + PA_{\Delta Cl}^T)\sin\theta & (A_{\Delta Cl}P - PA_{\Delta Cl}^T)\cos\theta \\ (PA_{\Delta Cl}^T - A_{\Delta Cl}P)\cos\theta & (A_{\Delta Cl}P + PA_{\Delta Cl}^T)\sin\theta \end{bmatrix} = \begin{bmatrix} \Pi_{11} & \Pi_{12} \\ \Pi_{21} & \Pi_{22} \end{bmatrix} + sym\left\{\begin{bmatrix} \Delta_{11} & \Delta_{12} \\ \Delta_{21} & \Delta_{22} \end{bmatrix}\right\} < 0,$$

where

$$\Pi_{11} = \Pi_{22} = \begin{pmatrix} A_0 P_S + P_S A_0^T + B_0 D_C C P_S + P_S C^T D_C^T B^T & B_0 C_C P_C + P_S C^T B_C^T \\ B_C C P_S + P_C C_C^T B_0^T & A_C P_C + P_C A_C^T \end{pmatrix} \sin\theta,$$

$$\Pi_{12} = -\Pi_{21} = \begin{pmatrix} A_0 P_S - P_S A_0^T + B_0 D_C C P_S - P_S C^T D_C^T B^T & B_0 C_C P_C - P_S C^T B_C^T \\ B_C C P_S - P_C C_C^T B_0^T & A_C P_C - P_C A_C^T \end{pmatrix} \cos\theta, \tag{36}$$

$$\Delta_{11} = \Delta_{22} = \begin{pmatrix} M_A F_A R_A P_S + M_B F_B R_B D_C C P_S & M_B F_B R_B C_C P_C \\ 0 & 0 \end{pmatrix} \sin\theta,$$

$$\Delta_{12} = -\Delta_{21} = \begin{pmatrix} M_A F_A R_A P_S - M_B F_B R_B D_C C P_S & M_B F_B R_B C_C P_C \\ 0 & 0 \end{pmatrix} \cos\theta.$$

Applying Lemma 4 to the second part in the right side of (35), one has

$$Sym\left\{\begin{bmatrix} \Delta_{11} & \Delta_{12} \\ \Delta_{21} & \Delta_{22} \end{bmatrix}\right\} = Sym\left\{\begin{bmatrix} M_A\sin\theta & M_B\sin\theta & 0 & M_A\cos\theta & M_B\cos\theta & 0 \\ 0 & 0 & 0 & 0 & 0 & 0 \\ -M_A\cos\theta & -M_B\cos\theta & 0 & M_A\sin\theta & M_B\sin\theta & 0 \\ 0 & 0 & 0 & 0 & 0 & 0 \end{bmatrix}\right.$$
$$\times \begin{bmatrix} F_A & 0 & 0 & 0 & 0 & 0 \\ 0 & F_B & 0 & 0 & 0 & 0 \\ 0 & 0 & 0 & 0 & 0 & 0 \\ 0 & 0 & 0 & F_A & 0 & 0 \\ 0 & 0 & 0 & 0 & F_B & 0 \\ 0 & 0 & 0 & 0 & 0 & 0 \end{bmatrix} \times \begin{bmatrix} R_A P_S & 0 & 0 & 0 \\ R_B D_C C P_S & R_B C_C P_C & 0 & 0 \\ 0 & 0 & 0 & 0 \\ 0 & 0 & R_A P_S & 0 \\ 0 & 0 & R_B D_C C P_S & R_B C_C P_C \\ 0 & 0 & 0 & 0 \end{bmatrix}\right\}$$

$$\leq \eta \begin{bmatrix} M_A\sin\theta & M_B\sin\theta & 0 & M_A\cos\theta & M_B\cos\theta & 0 \\ 0 & 0 & 0 & 0 & 0 & 0 \\ -M_A\cos\theta & -M_B\cos\theta & 0 & M_A\sin\theta & M_B\sin\theta & 0 \\ 0 & 0 & 0 & 0 & 0 & 0 \end{bmatrix} \times \tag{37}$$

$$\begin{bmatrix} M_A\sin\theta & M_B\sin\theta & 0 & M_A\cos\theta & M_B\cos\theta & 0 \\ 0 & 0 & 0 & 0 & 0 & 0 \\ -M_A\cos\theta & -M_B\cos\theta & 0 & M_A\sin\theta & M_B\sin\theta & 0 \\ 0 & 0 & 0 & 0 & 0 & 0 \end{bmatrix}^T$$

$$+\eta^{-1}\begin{bmatrix} R_A P_S & 0 & 0 & 0 \\ R_B D_C C P_S & R_B C_C P_C & 0 & 0 \\ 0 & 0 & 0 & 0 \\ 0 & 0 & R_A P_S & 0 \\ 0 & 0 & R_B D_C C P_S & R_B C_C P_C \\ 0 & 0 & 0 & 0 \end{bmatrix}^T \begin{bmatrix} R_A P_S & 0 & 0 & 0 \\ R_B D_C C P_S & R_B C_C P_C & 0 & 0 \\ 0 & 0 & 0 & 0 \\ 0 & 0 & R_A P_S & 0 \\ 0 & 0 & R_B D_C C P_S & R_B C_C P_C \\ 0 & 0 & 0 & 0 \end{bmatrix}.$$

Substituting (37) into (35) and using the Schur complement of (32) one has

$$\begin{bmatrix} \Sigma + \eta MM^T & R^T \\ \bullet & -\eta I \end{bmatrix} < 0, \quad (38)$$

in which

$$\hat{\Sigma} = \begin{bmatrix} \hat{\Sigma}_{11} & \hat{\Sigma}_{12} \\ \hat{\Sigma}_{21} & \hat{\Sigma}_{22} \end{bmatrix}, \hat{\Sigma}_{11} = \hat{\Sigma}_{22} = \begin{bmatrix} A_0 P_S + P_S A_0^T + B_0 D_C C P_S + P_S C^T D_C^T B^T & B_0 C_C P_C + P_S C^T B_C^T \\ B_C C P_S + P_C C_C^T B_0^T & A_C P_C + P_C A_C^T \end{bmatrix} \sin\theta, \quad (39)$$

$$\hat{\Sigma}_{12} = -\hat{\Sigma}_{21} = \begin{bmatrix} A_0 P_S - P_S A_0^T + B_0 D_C C P_S - P_S C^T D_C^T B^T & B_0 C_C P_C - P_S C^T B_C^T \\ B_C C P_S - P_C C_C^T B_0^T & A_C P_C - P_C A_C^T \end{bmatrix} \cos\theta.$$

The matrix inequality (38) is not linear because of various multiplications of variables. Accordingly, by changing variables as follows

$$T_1 = A_C P_C, T_2 = B_C C P_S, T_3 = C_C P_C, T_4 = D_C C P_S, \quad (40)$$

the matrix inequality (32) is obtained. ∎

**Corollary1**: Although Theorem 1 and Theorem 2 are allocated to robust stabilization of uncertain FO-LTI systems of form (1), the proposed method can be easily used for the case of certain systems by solving the LMI constraints $\Sigma < 0$ in these theorems, respectively.

**Proof:** The proof is straightforward by assuming $A_{\Delta Cl} = 0$ in proof procedure of Theorem 1 and Theorem 2.

## 4. Numerical examples

In this section, some numerical examples are given to demonstrate the applicability of the proposed method. In this paper, we use YALMIP parser [30] and SeDuMi [31] solver in Matlab tool [32] in order to assess the feasibility of the proposed constraints to obtain the controller parameters.

4.1.    Example 1 *for the* $0 < \alpha < 1$ *case*

Consider the dynamic output feedback stabilization problem of the uncertain fractional-order system (1) is considered with $\alpha = 0.75$ and $A \in A_I = [\underline{A}, \overline{A}], B \in B_I = [\underline{B}, \overline{B}]$, where

$$\underline{A} = \begin{bmatrix} 2 & -8 & 1 \\ 9 & 6 & 1 \\ 1 & 2 & -1 \end{bmatrix}, \overline{A} = \begin{bmatrix} 2.5 & -7 & 1.5 \\ 9.5 & 6.5 & 1.5 \\ 1.5 & 2.5 & -0.5 \end{bmatrix}, \underline{B} = \begin{bmatrix} 1 \\ -1 \\ 0 \end{bmatrix}, \overline{B} = \begin{bmatrix} 1.5 \\ -0.6 \\ 0 \end{bmatrix}, C = [1 \quad 0 \quad 1]. \quad (41)$$

The eigenvalues of $A$, $A_{Cl}$, and stability boundaries $\pm\alpha\pi/2$ are demonstrated in Fig. 1. According to Lemma 2 and Fig. 1, system (1) with parameters in (41) is unstable since some of the eigenvalues of A are located on the right side of boundaries. However, according to Theorem 1, it can be concluded that this uncertain fractional-order system is asymptotically stabilizable utilizing the obtained dynamic output feedback controllers of arbitrary orders in the form of (18), tabulated in Table 1. The eigenvalues of $A_{Cl}$ are located in stability region which is also obvious in Fig. 1.

**Table 1** Controller parameters obtained by Theorem 1.

| $n_C$ | $A_C$ | $B_C$ | $C_C$ | $D_C$ |
|---|---|---|---|---|
| 0 | 0 | 0 | 0 | $-24.86$ |
| 1 | $-5.55$ | $-0.43$ | $-1.25$ | $-26.55$ |
| 2 | $\begin{bmatrix} -3.80 & -0.00 \\ -0.35 & -3.73 \end{bmatrix}$ | $\begin{bmatrix} 0.32 \\ 0.31 \end{bmatrix}$ | $\begin{bmatrix} -1.05 \\ -1.08 \end{bmatrix}^T$ | $-25.22$ |
| 3 | $\begin{bmatrix} -3.46 & 0.00 & 0.00 \\ -0.45 & -3.50 & 0.00 \\ -0.38 & -0.45 & -3.48 \end{bmatrix}$ | $\begin{bmatrix} 0.17 \\ 0.25 \\ 0.24 \end{bmatrix}$ | $\begin{bmatrix} -1.13 \\ -1.30 \\ -1.23 \end{bmatrix}^T$ | $-26.44$ |

The time response of the resulted uncertain closed-loop FO-LTI system of form (20), via obtained controllers with $n_C = 0$ and 1 are illustrated in Fig. 2, where all the states asymptotically converge to zero. It can be concluded that, the obtained dynamic output feedback controllers, even with a low order of $n_C = 1$, have more appropriate stabilizing actions compared to static one. Furthermore, it is highly unrealistic to assume that all states are measurable. Accordingly, unlike state feedback procedure, all components of the state vector are not needed to be measured according to the matrix $C$.

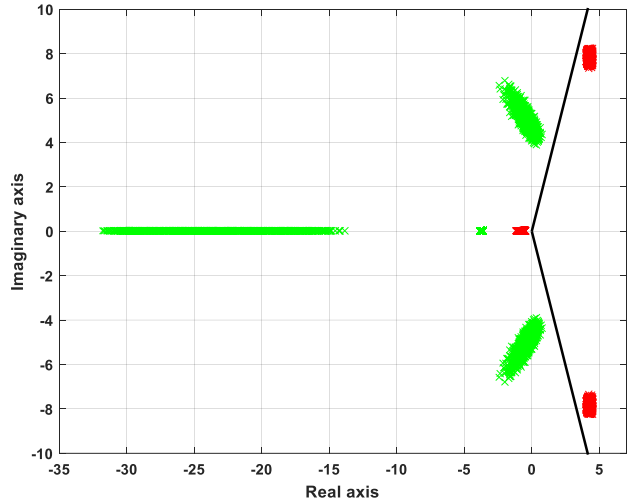

Fig. 1. The location of eigenvalues of the uncertain open-loop system (red) and closed-loop system via obtained output feedback controller with $n_c = 2$ (green) in Example 1.

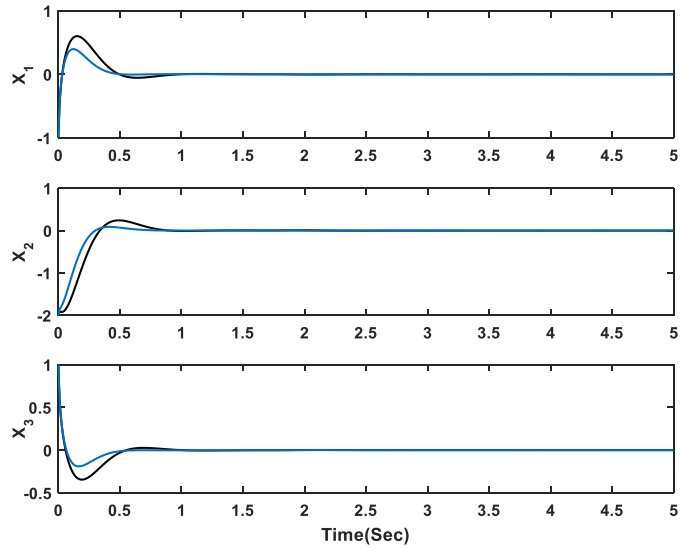

Fig. 2. The time response of the closed-loop system in Example 1 via obtained output feedback controllers with $n_c = 0$ (black) $n_c = 1$ (blue).

4.2.  Example *2 for the* $1 \leq \alpha < 2$ *case*

Dynamic output feedback stabilization problem of the interval fractional-order system (1) is considered with $\alpha = 1.2$ and $A \in A_I = [\underline{A}, \overline{A}]$, $B \in B_I = [\underline{B}, \overline{B}]$, where

$$\underline{A} = \begin{bmatrix} -1.1 & -1.5 & 3 \\ 0.8 & -2.6 & 0.7 \\ -1.4 & -4 & -1.2 \end{bmatrix}, \overline{A} = \begin{bmatrix} -0.9 & -1 & 4 \\ 1.2 & -2 & 1.3 \\ -1 & -3 & -0.8 \end{bmatrix}, \underline{B} = \begin{bmatrix} 1 \\ 1.9 \\ 0.9 \end{bmatrix}, \overline{B} = \begin{bmatrix} 1.1 \\ 2 \\ 1 \end{bmatrix}, C = [1 \quad 0 \quad -1]. \tag{42}$$

The eigenvalues of $A$, $A_{cl}$, and stability boundaries $\pm \alpha \pi/2$ are depicted in Fig. 3. According to Lemma 3 and Fig. 3, system (1) with parameters in (42) is unstable since some of the eigenvalues of $A$ are located on the right side of boundaries. However, according to Theorem 2, it can be concluded that this uncertain FO-LTI system is asymptotically stabilizable using the obtained dynamic output feedback controllers of arbitrary orders in the form of (18), presented in Table 2. The eigenvalues of $A_{cl}$ are located in stability region which is also obvious in Fig. 3.

Table 2 Controller parameters obtained by Theorem 2.

| $n_c$ | $A_c$ | $B_c$ | $C_c$ | $D_c$ |
|---|---|---|---|---|
| 0 | 0 | 0 | 0 | $-3.74$ |
| 1 | $-0.1453$ | $-0.0016$ | $-0.1198 \times 10^{-3}$ | $-1.41$ |
| 2 | $\begin{bmatrix} -0.1389 & -0.0005 \\ -0.0005 & -0.1389 \end{bmatrix}$ | $\begin{bmatrix} 0.0032 \\ 0.0042 \end{bmatrix}$ | $\begin{bmatrix} -0.23 \\ -0.24 \end{bmatrix}^T \times 10^{-3}$ | $-1.429$ |
| 3 | $\begin{bmatrix} -0.1334 & -0.0004 & -0.0006 \\ -0.0006 & -0.1332 & -0.0006 \\ -0.0005 & -0.0003 & -0.1333 \end{bmatrix}$ | $\begin{bmatrix} 0.0060 \\ 0.0038 \\ 0.0060 \end{bmatrix}$ | $\begin{bmatrix} -0.4079 \\ -0.2881 \\ -0.3544 \end{bmatrix}^T \times 10^{-3}$ | $-1.4236$ |

The time response of the resulted uncertain closed-loop FO-LTI system of form (20), via obtained controllers with $n_c = 0$, 1, and 3 are illustrated in Fig. 4, where all the states asymptotically converge to zero. It can be deduced that, the resulted dynamic output feedback controllers, even with a low order of $n_c = 1$, have more appropriate stabilizing actions compared to static one. Moreover, it is very unrealistic to suppose that all states are measurable. Accordingly, unlike state feedback procedure, all components of the state vector are not needed to be measured according to the matrix $C$.

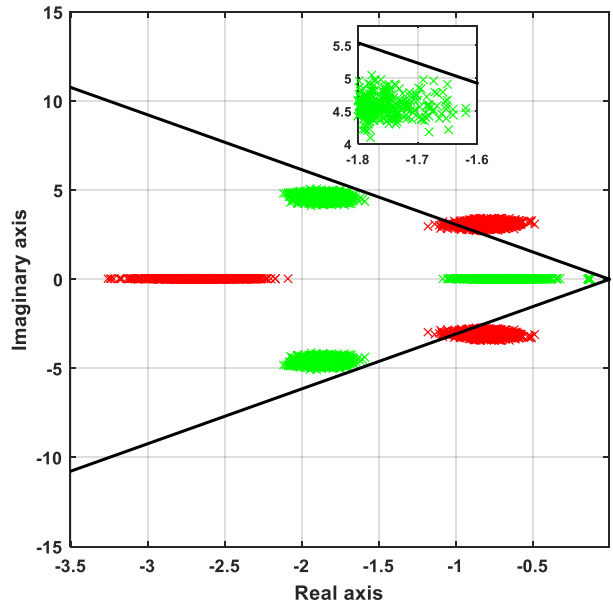

Fig. 3. The location of eigenvalues of the uncertain open-loop system (red) and closed-loop system via obtained output feedback controller with $n_c = 3$ (green) in Example 2.

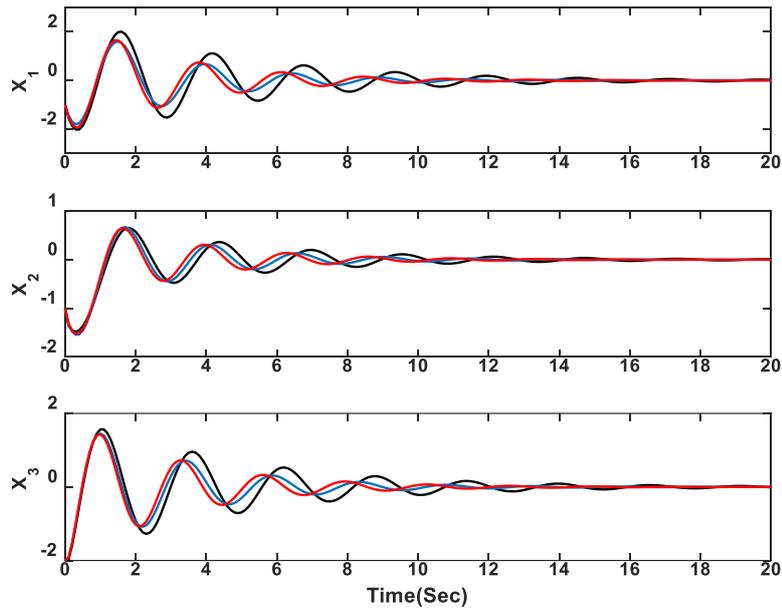

Fig. 4. The time response of the closed-loop system in Example 2 via obtained output feedback controllers with $n_c = 0$ (black), $n_c = 1$ (blue) $n_c = 3$ (red).

## 5. Conclusion

In this paper the problem of robust dynamic output stabilization of FO-LTI interval systems with the fractional order $0 < \alpha < 2$, in terms of LMIs is solved. Sufficient conditions are obtained for designing a stabilizing controller with a predetermined order, which can be chosen to be as low as possible for simpler implementation. Indeed by using proposed method, one can benefit from dynamic output feedback controller advantages with orders lower than the system order. The LMI-based procedures of developing robust stabilizing control are preserved in spite of the complexity of assuming the most complete model of linear controller, with direct feedthrough parameter. Eventually, two numerical examples have shown the effectiveness of our results.